\documentclass[a4paper,english,aps,prb,amsmath,showpacs,twocolumn,
floatfix]{revtex4}
\usepackage[T1]{fontenc}
\usepackage[latin1]{inputenc}
\usepackage{graphicx}
\makeatletter

\usepackage{graphicx}
\usepackage{graphicx}
\usepackage{graphicx}
\def\lsim{\mathrel {\vcenter {\baselineskip 0pt \kern 0pt
    \hbox{$<$} \kern 0pt \hbox{$\sim$} }}}
\usepackage{babel}
\usepackage{babel}
\usepackage{babel}
\makeatother
\begin{document}

\title{
Non-uniform liquid--crystalline phases of parallel hard rod-shaped particles:
From ellipsoids to cylinders.
}

\author{Y. Mart\'{i}nez--Rat\'{o}n}
\email{yuri@math.uc3m.es}
\affiliation{Grupo Interdisciplinar de Sistemas Complejos (GISC), 
Departamento de Matem\'{a}ticas,
Escuela Polit\'{e}cnica Superior, Universidad Carlos III de Madrid, 
Avenida de la Universidad 30, E--28911, Legan\'{e}s, Madrid, Spain}

\author{E. Velasco}
\email{enrique.velasco@uam.es}
\homepage{http://www.uam.es/enrique.velasco}
\affiliation{Departamento de F\'{\i}sica T\'{e}orica de la Materia Condensada
and Instituto de Ciencia de Materiales Nicol\'as Cabrera,
 Universidad Aut\'{o}noma de Madrid, E-28049 Madrid, Spain}

\date{\today}

\begin{abstract}
In this article we consider systems of parallel hard {\it superellipsoids},  
which can be viewed as a possible interpolation between ellipsoids of 
revolution and cylinders. Superellipsoids are characterized by an aspect 
ratio and an exponent $\alpha$ (shape parameter) which takes care 
of the geometry, with $\alpha=1$ corresponding to ellipsoids of revolution, 
while $\alpha=\infty$ is the limit
of cylinders. It is well known that, while hard parallel cylinders exhibit 
nematic, smectic, and solid phases, hard parallel
ellipsoids do not stabilize the smectic
phase, the nematic phase transforming directly into a solid as density is 
increased. We use computer simulation to find evidence
that for $\alpha\ge\alpha_c$, where $\alpha_c$ is a critical value
which the simulations estimate to be in the interval $1.2$--$1.3$, 
the smectic phase is stabilized. This is surprisingly
close to the ellipsoidal case. In addition, we use a density-functional 
approach, based on the Parsons--Lee approximation, to describe smectic
and columnar ordering. In combination with a free--volume theory for
the crystalline phase, a theoretical phase diagram is predicted.
While some qualitative features, such as the enhancement of smectic
stability for increasing $\alpha$, and the probable absence of a stable 
columnar phase, are correct, the precise location of
coexistence densities are quantitatively incorrect. 
\end{abstract}


\maketitle

\clearpage

\section{Introduction }
Hard interaction models have played an important role in the 
understanding of the nature and structure of simple liquids and crystals
made of particles with spherical symmetry. For anisotropic particles,
the hard ellipsoid (HE) and hard-spherocylinder (HSC) models have played
a role similar to that of the hard-sphere model (HS) in simple liquids, though
these models are not so universal as the HS model. In particular, HE and HSC
fluids exhibit an isotropic--nematic phase transition 
but, while the HSC fluid
shows a stable smectic phase, all evidence to date suggests that 
the HE fluid does not \cite{Frenkel}.
Clearly, the formation of the smectic phase must be 
the result of delicate packing effects directly related to particle shape.
This feature of the HE fluid is a clear disadvantage 
in the formulation of perturbation theories for liquid crystals 
\cite{Mederos}.

The problem of why hard ellipsoids do not get stabilised into a layered
smectic structure is in intriguing one. The properties of a perfectly 
aligned fluid of
ellipsoids can be mapped onto those of hard spheres, and the HS
fluid does not exhibit phases with order intermediate between the
fluid and the crystal. On top of that, orientational freedom probably 
plays against smectic formation. Recent simulation work on parallel
hard ellipsoids augmented by an isotropic square well have shown that,
in this system, smectic layers can be stabilised\cite{deMiguel}; 
however, in view of the rather artificial model potential used,
the result probably does not reflect any essentially interesting 
underlying property of ellipsoids. 

The physical reason for the absence of smectic order in the HE
fluid obviously lies in the geometrical properties of an
ellipsoid. Wen and Meyer\cite{Wen} addressed the general problem
of smectic formation in fluids of parallel hard rods. They provided
an explanation in terms of the entropy gain involved in the increased
packing efficiency in smectic layers, which more than outweighs the 
entropy loss associated with the onset of layering with respect to
the nematic phase. This efficiency is very much reduced in ellipsoids
due to particle shape, since ellipsoids arranged in a layer leave too
much void space. Alternatively one may think that increasing packing,
which would involve filling this space, entails interlocking between
the layers, which promotes crystalline order but discourages formation
of the smectic phase.

A question that can be asked to understand this problem from
a different perspective is the following: if we perturb
the shape of an ellipsoid in the direction of a spherocylinder or a
cylinder, both of which exhibit smectic phases \cite{Veerman}, 
for which particle shape does 
smectic stability set in? The answer to this question may provide 
some further insight into the relation between particle geometry and smectic
stability. As a bonus, it would help formulate more useful hard--body models 
that can be used in perturbation theories.

Only a few previous studies have addressed this issue. Of particular
relevance to our study is that of Evans\cite{Evans}, who used an Onsager 
second-virial coefficient
approximation to investigate smectic formation in fluids made of 
hard ellipsoids, hard spherocylinders and hard ellipocylinders, a particle
with a shape somehow intermediate between that of the first two. 
The ellipsoids, both parallel and with unconstrained orientations,
did not form a smectic phase before the crystal, while the other particles 
did at some particular density prior to crystallization. It was
concluded that ellipsoids are pathological in that they do not form
a smectic phase\cite{Evans}. 

In the present paper we again address this problem
by studying a {\it continuum} of particle shapes, in the limit of parallel
particles, but this time interpolating 
between the ellipsoid and the {\it cylinder} 
by means of a model, the {\it hard superellipsoid} (HSE) of revolution, 
containing an exponent $\alpha$ that can be varied continuously. 
Monte Carlo (MC) simulations are used to analyse the stability of the 
smectic phase and other phases with partial positional order (i.e. 
columnar phase) of the parallel model (PHSE), in the region of 
geometries close to the ellipsoid, and an approximate 
phase diagram as a function of $\alpha$ is obtained. The columnar phase
is not stable in the phase diagram. MC simulations of freely 
rotating cut spheres have shown that the columnar phase can be stabilised for 
some range of aspect ratios \cite{Frenkel2}. However simulations 
of parallel cylinders do not give conclusive evidence for the presence of 
columnar symmetry in the phase diagram \cite{Veerman}, and a recent 
fundamental measure theory (FMT) for parallel cylinders \cite{FMT}
also rules out the columnar phase as a thermodynamically stable phase.
However, one of the conclusions of the present paper is that 
the smectic phase {\it can} be stabilised with respect to the nematic
phase for some type of PHSE particles. Therefore, in the approximation 
of parallel particles, ellipsoids do not seem to be pathological; rather 
it is a continuum of particle
shapes, close to the ellipsoidal, that do not exhibit smectic order.
To complement these studies, a density--functional theory based on the 
Parsons--Lee approach \cite{Parsons}
was used to investigate the relative stability 
of the smectic phase with respect to other phases with lower symmetry.
For the crystal phase, a free--volume approximation is used.
The main conclusion that can be drawn is that the Parsons--Lee 
approximation, in combination with free--volume theory, only gives a
qualitative picture of the relative stability of the phases. The precise
location of the transition points are incorrect. Other features, such as the 
fact that the columnar phase is not stable, appear to be correctly 
described. Resort to other, more sophisticated theories, e.g. of the FMT type, 
would be needed to obtain a better description. However, 
although the initial framework of the FMT for general convex particles 
was initially developed by Rosenfeld \cite{Rosenfeld}, its application 
only works for isotropic fluids, as it was shown in Ref. \cite{Perera}.
For some simple geometries (such as parallelepipedic) the FMT  
can be formulated from first principles with restricted particle
orientations \cite{Cuesta}. The generalisation of the FMT to geometries 
like the ones described here seems to be a rather difficult task.

\section{Particle model}

\begin{figure}[h]
{\centering \resizebox*{9.0cm}{!}{\includegraphics{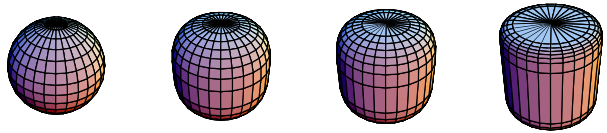}} \par}
\caption{\small Superellipsoids with $a=b$ (superspheres) for values 
of $\alpha$ equal to 1, 1.5, 2 and 5 (from left to right).} 
\label{fig1}
\end{figure}

We consider parallel particles with symmetry of revolution (axially 
symmetric about the $z$ axis). The superellipsoid results by making
a superellipse in the $x$--$z$ plane revolve around the $z$ axis, 
with the equation
\begin{eqnarray}
\left(\frac{R}{a}\right)^{2\alpha}+
\left|\frac{z}{b}\right|^{2\alpha}=1,\hspace{0.6cm}
R=\sqrt{x^2+y^2}. 
\label{1}
\end{eqnarray}
$a$ and $b$ are the semi-lengths of the particle in the $xy$ plane and
along the $z$ axis, respectively, and $\alpha$ is the shape exponent. This equation defines a 
particle with a geometry interpolating between the
ellipsoid ($\alpha=1$) and the cylinder ($\alpha=\infty$).
Note that the PHSE model is
different from that used by Evans\cite{Evans}, whose model
interpolates between an ellipsoid and a spherocylinder. 
Fig. \ref{fig1} depicts a few bodies (with $a=b$, which could be
called {\it superspheres}) for different values of
$\alpha$. Thermodynamic and structural properties of such a fluid of parallel
particles scale trivially with
particle elongation, so that it is sufficient to consider the case $a=b
=\sigma_0/2$, with $\sigma_0$ the particle breadth.

The stable low--density phase of the PHSE model is a nematic phase
(all particles oriented along the $z$ axis and with their centers of mass
disordered positionally). At high densities we expect a solid phase with
some crystalline structure (the question of which
structure is stabilised will be addressed later). At intermediate
densities we could, in principle, expect different phases with partial 
order (smectic or columnar) to get stabilised, depending on the value 
of $\alpha$.

\section{Theoretical tools}
\label{TT}

To explore the phase behaviour of the PHSE model we used constant--pressure
MC simulation. We have simulated systems of $N=1$--$2\times 10^3$ 
particles, with $\alpha=1.0$--$2.0$.
Both compression runs from the low-density nematic phase and
expansion runs from a high-density crystalline structure
were conducted. From these, equations of state are obtained, and 
the different phases may be identified. Relative stability
between nematic, smectic and crystal phases (some of which, as we will see, undergo
a first--order phase transition) cannot be ascertained,
since no attempt was made at computing absolute free energies.
Structural quantities, such as density profiles along the ordering 
direction, $\rho(z)$, and radial distribution functions parallel and 
perpendicular to this direction, $g_{\parallel}(z)$ and $g_{\perp}(R)$,
were also calculated. All this information allows us to get a picture
of the trends in phase behaviour that can be expected as the 
exponent $\alpha$ is varied.

To complement simulation results, a number of theoretical
analyses have been carried out. In the high--density region, 
the classical free--volume theory has been implemented.
The free--volume
free--energy density is written as
\begin{eqnarray}
\beta{\cal F}=-N\log{\left(\frac{v_{\hbox{\tiny free}}}{\Lambda^3}\right)}
\end{eqnarray}
where $N$ is the number of particles, $v_{\hbox{\tiny free}}$ the free volume 
available to a particle
when the others are kept fixed in their lattice positions, $\beta=1/kT$
and $\Lambda$ is the thermal wavelength. 
For parallel
hard cylinders this is an analytical function of density. In the general
case this is probably not true, and we have computed $v_{\hbox{\tiny free}}$
by MC integration. In the low--density region, a simple
virial expansion for the excess free energy,
\begin{eqnarray}
\beta{\cal F}_{\rm{ex}}[\rho]=&-&\frac{1}{2}\int d{\bf r}\int d{\bf r}'
\rho({\bf r})\rho({\bf r}') f({\bf r}-{\bf r}')\nonumber\\&&\nonumber\\&-&
\frac{1}{6}\int d{\bf r}\int d{\bf r}'
\int d{\bf r}^{\prime\prime}
\rho({\bf r})\rho({\bf r}')\rho({\bf r}^{\prime\prime})\nonumber\\ 
&\times&f({\bf r}-{\bf r}^{\prime}) f({\bf r}^{\prime}-{\bf r}^{\prime\prime})
f({\bf r}-{\bf r}^{\prime\prime})+...
\label{VI}
\end{eqnarray}
has been considered to calculate the nematic--smectic bifurcation line
\cite{Mulder}, with an aim to comparing with the Parson--Lee approach (see 
later). In the above expansion $f({\bf r})$ is the Mayer function of
two parallel PHSE, and $\rho({\bf r})$ the local density distribution. 
This expansion is meaningful only 
for the low--density nematic phase and expected to rapidly fail as
the system density is increased, but at least it may give an indication as to
whether the system is prone to developing smectic ordering 
and, if so, how this tendency depends on particle shape. 
Specifically, we have applied a bifurcation analysis based on the
above expansion (presented in Appendix A), using second-- and
third--order terms in $\rho({\bf r})$.
At higher density, i.e. in regions where the smectic or
columnar phases may be stable, an alternative is to use a 
Parsons--Lee (PL) scheme \cite{Parsons}. 
In the PL approach, an approximate resummation
of the virial series is performed, using the exact
second virial coefficient. That is, we take
\begin{eqnarray}
B_n^{\rm PHSE}=\frac{B_n^{\rm HS}}{B_2^{\rm HS}}B_2^{\rm PHSE}
\end{eqnarray}
where $B_n^{\rm HS}$ are hard--sphere virial coefficients.
With this scaling of the PHSE virial coefficients, the excess
free energy writes
\begin{eqnarray}
\beta{\cal F}_{\rm{ex}}[\rho]=-\frac{\Psi_{\rm HS}(\eta)}
{2B_2^{\rm HS}\rho_0}
\int d{\bf r}\int d{\bf r}'
\rho({\bf r})\rho({\bf r}') f({\bf r}-{\bf r}'),
\label{fex}
\end{eqnarray}
where $\Psi_{\rm HS}(\eta)$ is the excess free energy per particle
of a HS fluid of the same packing fraction $\eta$ as our fluid
of PHSE particles. The packing fraction is given by
$\eta=\rho_0v_0$, with $\rho_0$ the mean density and
$v_0$ the particle volume, given by
\begin{eqnarray}
v_0(\alpha)=
\frac{\pi\sigma_0^3}{12\alpha}B\left(\frac{1}{\alpha},\frac{1}{2\alpha}
\right),
\end{eqnarray}
where $B(x,y)$ is the beta function. For $\Psi_{\rm HS}(\eta)$ we 
use the Carnahan--Starling expression, $\Psi_{\rm HS}(\eta)=
(4-3\eta)\eta/(1-\eta)^2$. In the minimisations of the total free energy, 
\begin{eqnarray}
{\cal F}[\rho]&=&
\beta^{-1}\int d{\bf r}\rho({\bf r})\left\{\log{\left[\rho({\bf r})
\Lambda^3\right]}-1\right\}\nonumber\\
&+&{\cal F}_{\rm{ex}}[\rho],
\label{fid}
\end{eqnarray}
where ${\cal F}_{\rm{ex}}[\rho]$ is given by (\ref{VI}) or (\ref{fex}), the 
density profile $\rho({\bf r})$ is parametrised in some convenient way. For the
smectic and columnar phases at low densities 
 we use a truncated Fourier expansion. For high densities 
a Gaussian parameterization is used. 
Details of these calculations are given in Appendix B.

\section{Results}

\begin{figure}[h]
\hspace*{-0.5cm}
{\centering \resizebox*{9.5cm}{!}{\includegraphics{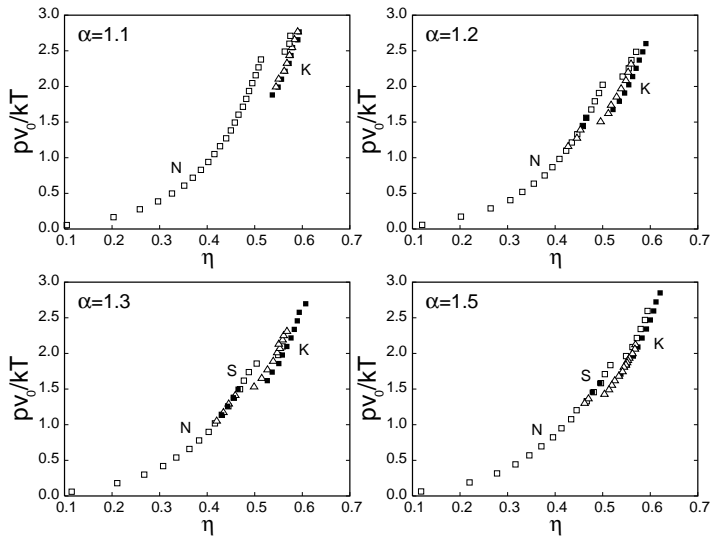}} \par}
\caption{\small Equation of state pressure--packing fraction for
various values of shape parameter $\alpha$ (shown in each panel),
as obtained from constant--pressure Monte Carlo simulations.
Open squares: compression runs starting from nematic phase.
Filled squares: expansion runs starting from crystalline phase
with ABC symmetry. Triangles: expansion runs starting from
columnar symmetry. Identification of phases is made with
symbols (N, S and K for nematic, smectic and crystal, respectively).}
\label{fig2}
\end{figure}

In this section we use the packing fraction $\eta$ as a 
convenient measure of density. For a given density, since $v_0$ is an 
increasing function
of $\alpha$, $\eta$ increases slightly from the ellipsoid to the cylinder.
In Fig. \ref{fig2}, MC data for reduced presure $pv_0/kT$ versus packing
fraction $\eta$, for all the particle shapes considered, are shown
(from now on we only show simulation results for the systems with
$N\simeq 10^3$ particles; selected checks with twice as many
particles did not give any quantitative differences).
In all cases the low--density phase is a nematic (N phase), since all
particles are parallel, whereas the system crystallises at high density
(K phase). We begin by discussing the high--density region, where the crystal
phase is the stable phase. 

As expected, compression runs starting 
from the low--density nematic phase (open squares in Fig. \ref{fig2})
ultimately create a defected crystal, via a first--order phase transition
(this appears as a clear density discontinuity in all 
cases). Expansion runs from a perfect high--density crystal (filled
squares in Fig. \ref{fig2}) give a solid branch with
defect--free structures in most instances. However, in small systems defects
consisting of columns of particles that have moved along the director
by a fraction of the unit cell are created. 
This may give the impression that averaged structures
correspond to a stable columnar phase.
A related situation was found in the simulations of Veerman and Frenkel
on parallel hard cylinders \cite{Veerman}, where a strong dependence 
of columnar stability on system size was found.

To investigate this, we prepared initial configurations with columnar
symmetry at high density; in expansion runs,
the system always stayed as a columnar phase (triangles in Fig. \ref{fig2}).
Therefore, for large system sizes, systems with columnar and crystalline 
structures
look as though they are actually mechanically stable, and there seems to be 
a large free--energy barrier between the two structures. However,
compressions from the nematic never give rise to configurations containing
partial columnar order (which would seem easier to generate than full 
three--dimensional order).
Without explicit calculation of free energies, no definite conclusion
can be reached on the relative stabilities of columnar and crystalline
phases; however, given that no clear evidence for columnar ordering has 
been found in simulations of parallel hard cylinders \cite{Veerman}, 
and that FMT calculations on the same system predict that columnar order is 
not thermodynamically stable \cite{FMT}, we believe it unlikely that a
columnar phase may get stabilised in PHSE.

\begin{figure}[h]
{\centering \resizebox*{6.0cm}{!}{\includegraphics{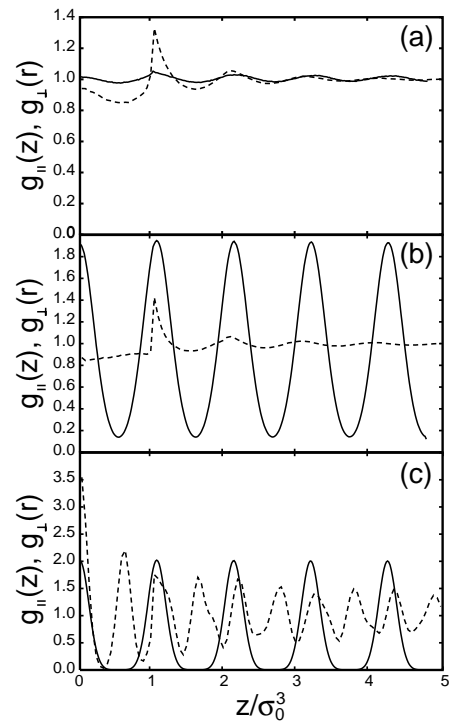}} \par}
\caption{\small Radial distribution functions along (continuous line)
and perpendicular to (dashed line) the 
director, $g_{\parallel}(z)$ and $g_{\perp}(R)$, for the PHSE fluid with
$\alpha=1.5$; (a) $\eta=0.38$; (b) $\eta=0.43$; and (c) $\eta=0.53$.}
\label{GRS}
\end{figure}

As the crystal is expanded to lower density the system  
looses translational order (either totally or partially), and becomes 
fluid via a discontinuity in density. Fig. \ref{GRS}(c) shows the
radial distribution functions along and perpendicular to the director,
$g_{\parallel}(z)$ and $g_{\perp}(R)$,
in the crystal phase for the case $\alpha=1.5$. 
Both present a high degree of structure, as expected in a crystalline
phase. Therefore all evidence suggests that there is a first--order 
fluid--crystal (i.e. freezing) transition. 

The question on the nature of the crystalline phase and how the
system is prepared in the expansion runs deserves some comments.
In fact the symmetry of the structure seems to change at some value of 
$\alpha$. The generic crystal structure consists of stacked triangular layers,
either in phase (AAA structure) or out of phase. The latter
may have ABCA..., ABAB..., or random--stacking structures, all
of these being compact structures (i.e. they share the same value of the
close--packing density). A priori one could consider the
AAA, ABC and ABAB structures to be the most stable candidates,
reflecting simple hexagonal--, face--centred--cubic-- and 
hexagonal--close--packed--like symmetries, 
respectively (the final structures obtained by the simulations
do not have these exact symmetries, since the intralayer
unit--cell distance does not exactly correspond to the interlayer
spacing expected in these structures; obviously this is a consequence
of the asymmetry of the particles along and perpendicular to their
symmetry axes). The situation, for lack of a more detailed
analysis based on free--energy estimations is, as usual,
uncertain (cf. the old debate on the hard--sphere crystal \cite{Frenkel3}),
presumably due to very small free--energy differences.
For freely rotating ellipsoids a recent 
work has shown the existence of a different crystal packing consisting 
of a simple monoclinic lattice with a basis of two ellipsoids 
with different orientations\cite{Schilling} (of course 
this structure cannot occur in our parallel--particle model).
In our case of parallel PHSE particles with $\alpha\le 1.5$, simulations with
between 8 to 12 layers along the $z$ direction may give 
structures with any type of stacking depending on how the
initial configuration is prepared (in the limit $\alpha=1$, we know that
ABC stacking is preferred \cite{Frenkel3}). However, in the limit $\alpha\to\infty$,
the AAA structure is clearly more stable, and it is expected that
somewhere in the interval $1.5<\alpha<\infty$ there is a change
in the nature of the crystal phase. Some preliminary simulations
indicate that the critical value may be between $1.5$ and $2.0$ (e.g. a
crystal with $\alpha=1.5$, when prepared with AAA stacking, shows a strong
tendency to evolve towards random stacking, whereas the system with $\alpha=2.0$,
when prepared with ABC stacking, evolves towards AAA stacking), but
free--volume theory gives a rather higher value (see later).

Now the central question of the present paper is whether there exists
an intermediate, stable fluid phase between the low--density nematic and the
high--density crystal phase. In the limit $\alpha=1$ (parallel ellipsoids)
ones knows for certain that there are no phases with partial (smectic
or columnar) order \cite{Frenkel}. By contrast, in the opposite limit 
$\alpha=\infty$ (parallel cylinders) previous MC simulations \cite{Veerman} 
indicate that the nematic phase changes to a smectic phase via a continuous
transition. Clearly in our interaction model, which interpolates between
these two limits, there must be a value of $\alpha$ beyond which the smectic phase
becomes stable. Actually this is the case. In Fig. \ref{fig2} cases where
a smectic phase has been identified are indicated by a corresponding label
(S for smectic). Evidence for the S phase
comes from density distributions (see later), as the pressure
shows no sign of a N--S transition, pointing to a continuous transition.
We should note that recent simulations on freely-rotating hard 
spherocylinders in the limit of infinite aspect ratio have shown that 
the nematic-smectic transition is of first order\cite{Polson}. Thus, we 
can conclude that the parallel alignment constraint is responsible for 
the second-order nature of the N-S transition of hard cylinders, and this
could also be the case in our model. By increasing the pressure further
(compression run), the smectic phase transforms into a defected crystal phase 
via a first--order phase transition. 

\begin{figure}[h]
{\centering \resizebox*{6.0cm}{!}{\includegraphics{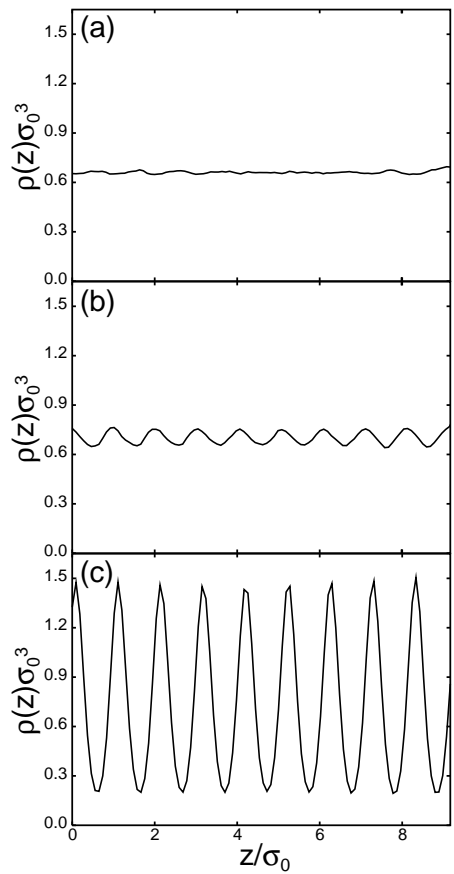}} \par}
\caption{\small Density profiles $\rho(z)$ for the PHSE fluid with
$\alpha=1.5$; (a) $\eta=0.39$; (b) $\eta=0.43$; and (c) $\eta=0.45$.}
\label{rhos}
\end{figure}

As mentioned above, our simulations indicate
that a smectic phase stabilises for {\it finite} $\alpha$, i.e. in the
range $\alpha_c\le\alpha<\infty$, where $\alpha_c$ is some critical value.
Indirect evidence comes from computation of
density profiles, smectic order parameter (not shown) and correlation functions. 
In Fig. \ref{rhos} the evolution of the density profile $\rho(z)$, as
density is increased, is shown for the case $\alpha=1.5$. The onset of smectic order 
is clearly identified by the smooth appearance of density peaks.
The density profiles exhibit a clear stratification at 
a packing fraction $\eta_{\hbox{\tiny NS}}\simeq 0.43$. 
That the high--density phase in question is smectic, and not crystalline,
can be concluded from 
the radial distribution functions along and perpendicular to the director,
shown in Figs. \ref{GRS}(a) and (b), 
which point to in--plane fluid--like correlations in the intermediate phase.

Analysis of the case $\alpha=1.2$, in contrast, suggests no evidence for
a smectic phase: the nematic fluid freezes directly 
into a crystal (see Fig. \ref{fig2}), similarly to the case of ellipsoids.
The intermediate case $\alpha=1.3$, however, does show signs of
smectic stability. 
>From all the information collected for the various systems analysed,
we estimate the critical value of $\alpha$ associated with the onset
of stability of the smectic phase to be in the range $\alpha_c=1.2$--$1.3$. 

\begin{figure}[h]
{\centering \resizebox*{8.0cm}{!}{\includegraphics{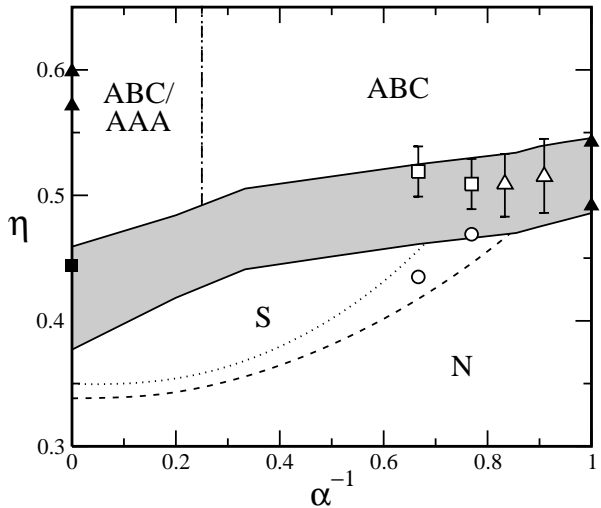}} \par}
\caption{\small Phase diagram in the plane packing fraction $\eta$--inverse
shape parameter $\alpha^{-1}$. Continuous lines: coexistence boundaries for
the smectic--solid transition from PL theory for the smectic and FV theory for the 
solid; shaded region: two--phase region from previous approximations;
dashed line: nematic-smectic spinodal from PL theory; dotted:
nematic-smectic spinodal from V3 theory; filled triangles: simulation results
for coexistence packing fractions of smectic--solid transition in parallel
cylinders ($\alpha=\infty$) from Veerman and Frenkel \cite{Veerman}, and for
liquid--solid coexistence in hard spheres \cite{Hoover} ($\alpha=1$);
filled square: nematic--smectic spinodal from computer simulation of 
Veerman and Frenkel \cite{Veerman}; open circles: our simulation estimates for
nematic--smectic spinodal; open squares: our simulation estimates for
first--order smectic--solid transition; open triangles: our simulation estimates for
first--order nematic--solid transition. In the latter two cases the vertical
bars only indicate limits of metastability. Labels indicate stable phases;
N, nematic; S, smectic; ABC and AAA, crystalline solids with corresponding 
symmetries. Vertical dot--dashed line: approximate limit of degeneracy of ABC and AAA
structures within FV theory.}
\label{fig3}
\end{figure}

Fig. \ref{fig3} summarises our results in the form of a phase diagram 
of packing fraction $\eta$ vs. inverse exponent $\alpha^{-1}$. 
For $\alpha\agt 1.2$ the smectic phase (S in the graph) 
is stable. Simulation data for the limit
$\alpha=\infty$ (cylinders, filled triangles and squares) are taken from 
simulations by Veerman and Frenkel \cite{Veerman},
and are essentially exact as they were inferred from simulations incorporating
free-energy calculations. Data for hard spheres (filled circles) are taken from Hoover and 
Ree \cite{Hoover}. In the other cases transition densities 
for the nematic--smectic, nematic--crystal and smectic--crystal transitions are 
only approximate. As already mentioned, the first (indicated by open circles)
is of second order. The others (open triangles and squares) are of first--order, 
with a wide density gap; since no free energies were computed,
we only plot, using vertical bars, approximate limits of metastability of the 
two phases involved  (the symbols are only the average packing--fraction values
inferred from the estimated limits of metastability). 

A second aim of our study is to rationalise the findings from computer
simulation using theoretical models. As mentioned in Sec. III, we have considered 
three theoretical models:
a low--density virial expansion up to third order in density for the
low--density nematic phase (V3), a resummed virial expansion of the 
Parsons--Lee type (PL) for intermediate densities, and free--volume
(FV) theory for the high--density crystal. 
The V3 theory was initially used to analyse possible bifurcations
of the nematic phase to smectic or columnar phases \cite{Mulder}. 
The qualitative
results as far as the continuous nematic--smectic transition is concerned 
are the same for both V3 and PL theories; 
they only differ quantitatively, as can be seen in Fig. \ref{fig3}. 
For large values of $\alpha^{-1}$ (the only available from simulations),
the MC data are bracketed by the two theories, but the trend that
the packing fraction at bifurcation increases with $\alpha^{-1}$ in this
regime is captured correctly. This increase is basically due to the 
decrease in particle volume as cylinders change to ellipsoids. The
differences between the V3 and PL predictions can be traced back
to their different treatment of correlations: V3 includes
three--body correlations exactly, but the remaining terms are neglected,
while PL only includes the exact two--body correlations but approximately
resums the higher--order density correlations.

The PL theory, together with the FV theory for the crystal, were used
to compute the nematic--solid and smectic--solid transitions.
These are first--order transitions, with a wide 
density gap. The two--phase region merges with the
nematic--smectic spinodal coming from lower densities at a critical end--point
located at $\alpha_c\simeq 1.2$; this value approximately
agrees with that inferred 
from the simulations. For lower values of $\alpha$ there is direct 
coexistence between nematic and solid phases, and the theory also seems
to agree with simulations in the limit of hard spheres (this agreement
is fortuitous as it is well known that the PL theory is inappropriate
to model spatial correlations in the hard-sphere fluid; the 
weighted-density theory \cite{Tarazona1} or a theory based
on FMT \cite{Tarazona2} are known to be more adequate).
In the opposite limit (cylinders), however, there is a big discrepancy with 
simulation, not only in the location of the nematic--smectic spinodal and
the smectic--solid phase transition, but also in the density gap of the latter 
which is significantly overestimated as compared to the MC simulations. 
In this limit the weighted-density type theories,
developed for a HSC fluid \cite{Somoza}, are obviously more appropriate to 
predict the spatial correlations. 
The PL and FV theories were also used to compute the smectic--columnar
and columnar--solid transitions. These results are not plotted in the
phase diagram since the region of columnar stability is always preempted
by crystallization directly from the smectic phase; this feature also
agrees with simulations which, as mentioned before, do not seem to give 
conclusive evidence for stable columnar ordering in this model.

\begin{figure}[h]
{\centering \resizebox*{8.0cm}{!}{\includegraphics{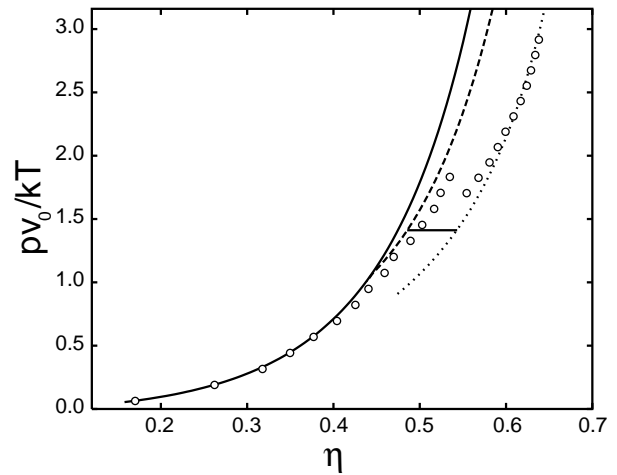}} \par}
\caption{\small Equation of state for the PHSE system with
$\alpha=1.5$. Symbols: MC results; continuous line: nematic branch as
obtained from CS approximation; dashed line: smectic branch from PL
theory; dotted line: solid branch according to FV theory.}
\label{EOS}
\end{figure}

The quality of the different approximations, in the density range where
each of them was used, can be checked by examining the equation of state (EOS).
This is done in Fig. \ref{EOS}, which refers to the case $\alpha=1.5$.
The figure also contains the MC simulation data. The PL theory, 
which reduces to the Carnahan--Starling approximation for the nematic phase
(continuous line), represents correctly the EOS up to packing fractions of 
$\eta\simeq 0.4$, just before the transition to the smectic phase occurs 
($\eta_{\hbox{\tiny NS}}\simeq 0.43$). 
The PL theory is not as accurate for the smectic phase
(dashed line), since it overestimates the pressure. The 
nematic--smectic transition density (bifurcation point) is reasonably well 
reproduced. In the solid branch (dotted line), the
equation of state is accurately represented (in fact better as the density 
approaches the close--packing limit), as expected, but the smectic--solid
transition is not correctly reproduced (value of pressure, location of
transition densities and density gap), due to the defects in the smectic
and solid equations of state.

\section{Conclusions}
We can conclude from our study that there is clear evidence that
in the PHSE fluid smecticity begins to be favoured entropically beyond
a critical value of the exponent $\alpha_c\simeq 1.2$--$1.3$;
this is surprisingly close to the value corresponding to ellipsoids
($\alpha=1$). Therefore, the fluid of parallel hard ellipsoids does not seem to be
pathological or special in not exhibiting a stable smectic phase;
rather, there is a family of particle shapes beyond the ellipsoid that do not
possess stable smectic phases, although the extent of the family in parameter space is small.

The explanation for this behaviour lies, of course, in the packing
efficiency of these hard--particle fluids. The
formation of the smectic phase is the result of a delicate packing
effect directly related to particle shape. In the HE fluid, the
ellipsoidal geometry creates a large tendency for particles to
interlock at their ends, which favors the stabilisation of the crystal phase
with detriment to the smectic phase. This effect is not relevant in
the case of HSC and sets in at some intermediate particle shape.

Our simulation results seem to indicate that the columnar phase is not stable 
for this family of parallel rods. But free-energy computations will be needed
to settle this question. However, in common with the case of parallel cylinders,
it is not likely that the columnar phase will be stable in our model.
All these conclusions correspond to the model
of parallel particles. Analysis should be extended to consider particles with
free orientational degrees of freedom. Given the high degree
of orientational order of the smectic phases, we do no think the conclusions
drawn from our study will be changed substantially.

As a byproduct of our analysis, we have assessed the validity of various
theoretical approximations by comparison with the simulation results.
A combination of free--volume and PL theories qualitatively
predicts the correct phase stability, but the agreement is far from 
quantitative, except in the parameter region close to the spheres; 
in particular, the value of
shape anisotropy beyond which smectic stability sets in is quite in
agreement with simulations. Use of the PL theory to
consistently describe all phases is not adequate, since this theory
progressively degrades as order builds up in the system:
therefore, the agreement is probably fortuitous.

An obvious avenue to improve the theoretical treatment is to use a more
sophisticated theory, which is still to be formulated. A promising approach
is FMT, which one hopes would correctly predict 
the relative stability of non-uniform phases; this belief is
based on FMT calculations applied to the particular case of parallel hard 
cylinders, as shown in Ref.  \cite{FMT}. However, the extension of this 
theory to superellipsoidal geometry would demand the use of strong 
approximations, in the line of Ref. \cite{Schmidt} where only 
the first terms of the excess free-energy asymptotic expansion
with respect to $\kappa^{-1}$ (with $\kappa$ the particle aspect ratio)
are taken into account. This procedure 
could disrupt the high predictive 
power of FMT as concerns the precise location of phase transitions 
and the relative stability of different non-uniform phases. Further 
investigation of this problem would be needed but is left for future work.

Interesting questions on the effect of particle shape on the 
depletion interaction between two anisotropic bodies mediated by 
small spherical particles can be studied within the present model. 
Recent studies have shown that particle geometry has 
a strong effect on the depletion forces \cite{Martinez-Raton}. The 
present model can be used to tune the geometry of the particle
with a single parameter, from ellipsoids to cylinders, and allows
the study of the evolution of the depletion potential with respect to 
$\alpha$ when two superellipsoids are immersed in a sea of small hard 
spheres. Work in this direction is in progress.    

\begin{acknowledgments}
Y.M.-R. gratefully acknowledges financial support from Ministerio de 
Educaci\'on y Ciencia (Spain) under a Ram\'on y Cajal research contract 
and the MOSAICO grant. This work is part of the research
projects Nos. FIS2005-05243-C02-01 and FIS2007-65869-C03-01,
also from Ministerio de Educaci\'on y Ciencia,
and grant No. S-0505/ESP-0299 from Comunidad Aut\'onoma de Madrid (Spain).
\end{acknowledgments}

\appendix

\section{Spinodal instability of N-S transition}
\label{BA}

The spinodal instability coincides with the location 
of the continuous N-S phase transition. The instability condition can be 
calculated by solving the following set of equations 
\begin{eqnarray}
1-\rho \hat{c}(\eta,q)=0,\quad \frac{\partial \hat{c}(\eta,q)}
{\partial q}=0,
\label{system}
\end{eqnarray}
where $\hat{c}(\eta,q)$ is the Fourier transform of the direct 
correlation function. The equations are to be solved for $\eta^*$ and $q^*$, 
which are the value of the packing fraction and the 
wave number of the smectic phase, both at bifurcation. The second equation
refers to the absolute minimum of $-\hat{c}(\eta,q)$ as this is 
a strongly oscillatory function of $q$. 

To calculate the spinodal curve a model for the direct correlation function
is needed. The third virial approach, Eqn. (\ref{VI}), gives
\begin{eqnarray}
\hat{c}(\eta,q)=\hat{f}(q)+\rho\Delta(q),
\end{eqnarray}
where the Fourier transform of the Mayer function 
$\hat{f}(q)=\int d{\bf r}e^{iqz}f({\bf r})$ gives the 
following result for a PHSE particle with unit breadth and height:
\begin{eqnarray}
-\hat{f}(q)=\frac{4\pi}{q}\int_0^1dz z \sin\left[q \left(1-z^{2\alpha}
\right)^{1/(2\alpha)}\right],
\end{eqnarray}
and 
\begin{eqnarray}
-\Delta(q)&=&\int d{\bf r}e^{iqz}f({\bf r})
\Delta V({\bf r}),\\
\Delta V({\bf r})&=&\int d{\bf r}'f({\bf r}')f({\bf r}-{\bf r}').
\end{eqnarray}
For each value of $q$ (which is understood to be given in units of $\sigma_0$), 
taken from a set of equally-spaced points, 
the function $\Delta(q)$ was calculated by MC integration
for a fixed $\alpha$. Then the equations 
(\ref{system}) were solved to find the spinodal curve $\eta^*(\alpha)$. 
Finally, the Parsons-Lee approach gives 
${\hat{c}(\eta,q)=\frac{1}{4}\Psi_{\rm{HS}}(\eta)\hat{f}(q)}$.

\section{Parameterizations of the Parsons-Lee theory}
\label{PL}

The density distributions for smectic and columnar symmetries were 
parameterized using two different approaches. The first, based on 
a Fourier expansion, is more adequate for relatively low mean densities,
while the other, a parameterization based on two parameters (one being
a measure of the width of the density peaks, the other being
the smectic or columnar lattice parameters --periods), 
is more appropriate for high--density phases, as the numerical convergence 
of the minimization schemes is much easier when the parameter space is reduced 
to two variables. 
We have used the following parameterization for the density profile of 
the smectic phase:
\begin{eqnarray}
\rho(z)=\frac{\rho_0}{I_0(\lambda)}\exp{\left(\lambda \cos qz\right)},
\label{uno}
\end{eqnarray}
where $\rho_0=d^{-1}\int_0^d dz\rho(z)$ is the mean density,
$q=2\pi/d$ the wave number ($d$ being the smectic period),
$\lambda$ the minimization parameter, and $I_0(\lambda)$ the zeroth
order modified Bessel function. Inserting this expression into the 
Parsons--Lee functional, Eqn. (\ref{fex}), we find
the following expression for $\varphi_{\rm{ex}}=\beta{\cal F}_{\rm ex}/N$, 
the excess part of the free-energy per particle and unit thermal energy:
\begin{eqnarray}
\varphi_{\rm{ex}}=-\frac{\Psi_{\rm{HS}}(\eta)}{2B_2^{\rm{HS}}\rho_0^2}d^{-1}
\int_0^ddz\rho(z)\int_{-\infty}^{\infty} dz' \rho(z')\tilde{f}
(z-z'), \nonumber\\
\label{varphi}
\end{eqnarray}
where $B_2^{\rm{HS}}=4v_0(\alpha)$ (since the volume of the reference 
HS particle is made to coincide with that of the PHSE). Also we have defined 
$\tilde{f}(z)=\int d{\bf r}_{\perp}f({\bf r}_{\perp})$ as the integrated 
Mayer function over the transverse area, with ${\bf r}_{\perp}=(x,y)$. 
Note that the HS 
free-energy per particle is $\Psi_{\rm{HS}}(\eta)=(4-3\eta)\eta/(1-\eta)^2$. 
By calculating the integral involved in Eq. (\ref{varphi}) we find explicitly
\begin{eqnarray}
 \varphi_{\rm{ex}}=2\Psi_{\rm{HS}}(\eta)C_{\alpha}
\chi(\lambda;\alpha),
\end{eqnarray}
where 
\begin{eqnarray}
C_{\alpha}&=&\frac{3\alpha}{2}B^{-1}\left(\frac{1}{\alpha},\frac{1}{2\alpha}
\right),\\
\chi(\lambda;\alpha)&=&I_0^{-2}(\lambda)\int_0^1 dz (1-z^{2\alpha})^{1/\alpha}
I_0\left(2\lambda\cos \frac{q^*z}{2}\right),\nonumber\\
\end{eqnarray}
and $q^*=2\pi\sigma_0/d$. Finally, the ideal part of the free-energy 
per particle and unit thermal energy, $\varphi_{\rm{id}}\equiv\beta
{\cal F}_{\rm id}/N$, can be found as 
\begin{eqnarray}
\varphi_{\rm{id}}= \ln \eta-1+\lambda\frac{I_1(\lambda)}
{I_0(\lambda)}-\ln I_0(\lambda),
\end{eqnarray}
where $I_1(x)$ is the first order modified Bessel function. We have 
minimized the total energy $\varphi=\varphi_{\rm{id}}+\varphi_{\rm{ex}}$ 
with respect to $\lambda$ and $d$ for a fixed value of $\eta$.

For the columnar phase, the parameterization chosen is
a sum of Gaussian peaks centered at the sites of the triangular lattice:
\begin{eqnarray}
\rho({\bf r}_{\perp})=\frac{\lambda_0\rho_0A_{\rm{cell}}}{\pi}\sum_{\bf k}
\exp{\left[-\lambda_0\left({\bf r}_{\perp}-{\bf R}_{\bf k}\right)^2\right]},
\end{eqnarray} 
and normalized in such a way that integration over the unit cell of area 
$A_{\rm{cell}}=\sqrt{3}a^2/2$ ($a$ being the lattice parameter of 
the triangular cell) gives the mean density $\rho_0$.
The position of the lattice sites are 
${\bf R}_{\bf k}=k_1{\bf a}_1+k_2{\bf a}_2$ ($k_i\in Z$) with 
$\displaystyle{{\bf a}_n=\frac{a}{2}\left(\sqrt{3},(-1)^n\right)}$ 
being the vectors of the triangular lattice. The 
Gaussian width is controlled by the parameter $\lambda_0$ which, 
together with $a$, define the set of minimization variables.
The excess part of the free-energy per particle can be calculated as
\begin{eqnarray}
\varphi_{\rm{ex}}&=&
\frac{\gamma}{\pi}A_{\rm{cell}}^*\Psi_{\rm{HS}}(\eta)C_{\alpha} 
\sum_{{\bf k}}e^{-\frac{\gamma}{2} \left(R_{\bf k}^*\right)^2}\nonumber\\
&\times&\int_0^{1/a^*}dr r 
\left[1-\left(ra^*\right)^{2\alpha}\right]^{1/(2\alpha)} 
e^{-\frac{\gamma}{2} r^2}I_0(\gamma R_{\bf k}^* r),\nonumber\\
\label{lala}
\end{eqnarray}
where $\gamma=\lambda_0 a^2$, $a^*=a/\sigma_0$, $R_{\bf k}^*=
|{\bf R}_{\bf k}|/a$, and $\displaystyle{A_{\rm{cell}}^*=
\frac{\sqrt{3}}{2}\left(a^*\right)^2}$ 
(the unit cell area in dimensionless 
units). Correspondingly 
the ideal part of the free-energy per particle is defined by
\begin{eqnarray}
\varphi_{\rm{id}}&=&\ln\left[\sqrt{3}\eta \gamma/(2\pi)\right] -1+ 
\frac{1}{\pi}\int_0^{2\pi}d\phi\int_0^{\infty}dr r \nonumber\\
&\times&e^{-r^2}\ln \left[
\sum_{\bf k}e^{-\left({\bf r}-\sqrt{\gamma}{\bf R}_{\bf k}^*\right)^2}\right] 
\label{useful}
\end{eqnarray}
An useful approximation for the ideal part when $\gamma\gg 1$ 
can be obtained by taking ${\bf R}_{\bf k}^*\to {\bf 0}$ in Eqn. 
(\ref{useful}), with the result
\begin{eqnarray}
\varphi_{\rm{id}}=\ln\left[\sqrt{3}\eta\gamma/(2\pi)\right]-2.
\end{eqnarray}
Also, the excess part (\ref{lala}) can be approximated by 
taking only the terms ${\bf k}=(0,\pm 1)$, ${\bf k}=(\pm 1,0)$,
${\bf k}=(1,-1)$ and ${\bf k}=(-1,1)$ 
in the sum (i.e. considering only the nearest neighbours of a given site).

Now we proceed to describe the other parametrization used for minimization 
at low densities: a truncated Fourier expansion of the density 
the density profile.
This expansion for the smectic symmetry is 
\begin{eqnarray}
\rho(z)=\rho_0\psi(z)=\rho_0\left[1+\sum_{n>0}^N\rho_n\cos(qnz)\right],
\end{eqnarray}
with $\rho_n$ the Fourier amplitudes. Using this parameterization,
the excess part of the free-energy per particle reads
\begin{eqnarray}
\varphi_{\rm{ex}}=\Psi_{\rm{HS}}(\eta)\sum_{n\geq 0}^N\theta_nT_n\rho_n^2,
\end{eqnarray}
where $\theta_n=(1+\delta_{n,0})/2$ and 
\begin{eqnarray}
T_n=\frac{4C_{\alpha}}{q_n}
\int_0^1dz z 
\sin\left[q_n\left(1-z^{2\alpha}\right)^{1/(2\alpha)}\right],
\end{eqnarray}
with $q_n=q\sigma_0n$, while the ideal part is
\begin{eqnarray}
\varphi_{\rm{id}}=\ln \eta-1+2\int_0^{1/2}dz \psi(dz)\ln\psi(dz).
\end{eqnarray}

For the columnar symmetry the Fourier expansion of the density distribution
reads
\begin{eqnarray}
\rho({\bf r}_{\perp})&=&\rho_0\psi({\bf r}_{\perp})=
\rho_0\Bigg[1+\sum_{n_1,n_2>0}^{N_1,N_2}\rho_{n_1n_2}\nonumber\\
&\times& \cos\left(\frac{2\pi n_1x}{
a\sqrt{3}}\right)\cos\left(\frac{2\pi n_2y}{a}\right)\Bigg],
\label{col}
\end{eqnarray}
where $\rho_{n_1n_2}$ are the two-dimensional 
Fourier amplitudes.  
After insertion of Eqn. (\ref{col}) into the Parsons-Lee expression for the 
excess free energy, we obtain
\begin{eqnarray}
\varphi_{\rm{ex}}=\Psi_{\rm{HS}}(\eta)\sum_{n1,n2\geq 0}^{N_1,N_2}\theta_{n_1n_2}
T_{n_1n_2}\rho_{n_1n_2}^2,
\end{eqnarray}
for the excess free-energy per particle,
where $\theta_{n_1n_2}=\left(\delta_{n_1+n_2,0}+\delta_{n_1,0}+\delta_{n_2,0}
+1\right)/4$, and
\begin{eqnarray}
T_{n_1n_2}=4C_{\alpha}
\int_0^1
dz z \left(1-z^{2\alpha}\right)^{1/(2\alpha)}J_0(q_{n_1n_2}z),
\end{eqnarray}
with $J_0(x)$ the zeroth order Bessel function and 
$\displaystyle{q_{n_1n_2}=\frac{2\pi}{a^*}\sqrt{\frac{n_1^2}{3}+n_2^2}}$.
The ideal part of the free-energy per particle for columnar symmetry 
can be calculated as 
\begin{eqnarray}
\varphi_{\rm{id}}&=&\ln \eta-1+\frac{8}{3\sqrt{3}a^2}
\int_0^{\sqrt{3}a/2}dx\int_0^{a-x/\sqrt{3}}dy\nonumber\\
&\times&\psi(x,y)\ln\psi(x,y).
\end{eqnarray}
This time the total free-energy per particle should be 
minimized with respect to the lattice parameter $a$ and also with respect 
to the 
Fourier amplitudes $\rho_{n_1n_2}$. While the Fourier parameterizations 
were used for the minimization of the total free-energy of 
the smectic and columnar phases at low densities (up to $\eta \sim 0.55$), 
the corresponding 
parameterizations  using only two parameters, described above,
were used to minimize the free energies per particle of these phases 
at higher densities. Both approaches were checked for consistency
at intermediate densities.

\end{document}